\pdfoutput=1
\RequirePackage{ifpdf}
\ifpdf 
\documentclass[pdftex]{sigma}
\else
\documentclass{sigma}
\fi

\numberwithin{equation}{section}

\newcommand{\ket}[1]{\rvert#1\rangle}

\newcommand{\braket}[2]{\langle #1\rvert#2\rangle}

\begin{document}

\newcommand{\arXivNumber}{1705.04841}

\renewcommand{\thefootnote}{}

\renewcommand{\PaperNumber}{074}

\FirstPageHeading

\ShortArticleName{Coherent Transport in Photonic Lattices: A Survey of Recent Analytic Results}

\ArticleName{Coherent Transport in Photonic Lattices:\\ A Survey of Recent Analytic Results\footnote{This paper is a~contribution to the Special Issue on Symmetries and Integrability of Dif\/ference Equations. The full collection is available at \href{http://www.emis.de/journals/SIGMA/SIDE12.html}{http://www.emis.de/journals/SIGMA/SIDE12.html}}}

\Author{\'Eric-Olivier BOSS\'E and Luc VINET}

\AuthorNameForHeading{\'E.-O.~Boss\'e and L.~Vinet}

\Address{Centre de Recherches Math\'ematiques, Universit\'e de Montr\'eal,\\ C.P.~6128, Succ.~Centre-ville, Montr\'eal, QC, Canada, H3C 3J7}
\Email{\href{mailto:eric-olivier.bosse@umontreal.ca}{eric-olivier.bosse@umontreal.ca}, \href{mailto:luc.vinet@umontreal.ca}{luc.vinet@umontreal.ca}}

\ArticleDates{Received May 13, 2017, in f\/inal form September 09, 2017; Published online September 14, 2017}

\Abstract{The analytic specif\/ications of photonic lattices with fractional revival (FR) and perfect state transfer (PST) are reviewed. The approach to their design which is based on orthogonal polynomials is highlighted. A compendium of analytic models with PST is of\/fered. New results on their FR properties are included. The nearest-neighbour approximation is adopted in most of the review; one analytic example with next-to-nearest neighbour interactions is also presented.}

\Keywords{perfect state transfer; fractional revival; quantum information}

\Classification{33C45; 81P45; 81V80}

\renewcommand{\thefootnote}{\arabic{footnote}}
\setcounter{footnote}{0}

\section{Introduction}
This paper reviews recent analytic results on the perfect transfer and fractional revival of states or excitations in photonic lattices using a general method that also applies to spin chains, tight-binding Hamiltonians, etc. Controlling the evolution of quantum states in devices is of high relevance in areas such as quantum information. With an eye to of\/fering a clean basis for experimental testing, it is also quite practical to look for exact analytical descriptions of such systems. Another objective is to minimize the need for external controls to achieve given tasks, in this respect a key idea is to construct devices whose dynamics will yield on its own the desired ef\/fect leaving only the input/output operations as external interactions.

A basic task is the transportation of quantum states from one location to another. Ideally one wishes to have perfect state transfer (PST) whereby a quantum state at an initial position is found with probability $1$ at the f\/inal destination. While there has been seminal work on optical Bloch oscillations in waveguide arrays, see for instance \cite{corrielli2013fractional,Iwanow2005, peschel1998optical}, this review focuses on coherent transport in spin-inspired photonic lattices. Since it has been suggested that spin chains \cite{11} with properly modulated couplings \cite{4,12} could move a qubit from end-to-end with probability~$1$ in a given time, there has been enormous interest in this question and indeed various analytic models have been found \cite{7,6,17,16,14,15,3,2}. See \cite{Bose2007,Kay2010,13} for reviews. Remarkably experimental verif\/ications have been of\/fered \cite{9,10,18} for the simplest (Krawtchouk) model with PST by using arrays of evanescently coupled waveguides \cite{20,21, 19}. These can be pictured as a set of optical f\/ibers stacked side by side in a plane. The point is that in the nearest-neighbour approximation, the equations of coupled mode theory describing the propagation of a single photon in waveguide arrays are mathematically identical to those governing the one-excitation dynamics of a certain spin chain with non-uniform couplings. Each qubit is represented by a~single waveguide and the presence or absence of a photon in a waveguide corresponds to the~$\ket{1}$ or~$\ket{0}$ state of the qubit at the corresponding site. The evolution time $t$ of the spin chain is identif\/ied with the propagation distance $z$ along the waveguides. Therefore PST is tantamount to f\/inding in the last waveguide after a distance $Z_{\rm PST}$, the photon initially inserted (at $z=0$) in the f\/irst f\/iber of the array\footnote{Strictly speaking in the photonic lattice realizations, one is producing an excitation transfer. In a spin chain, a state, i.e., an arbitrary combination of spin up and down will be transported. This is modeled in \cite{10} in an optical settings by encoding a state in polarizations.}. The required couplings between the waveguides are engineered by adjusting the distances so that they match the theoretical values. The analysis that underscores the design of the spin chains/photonic lattices with PST shows that the couplings and local magnetic f\/ields/propagation constants correspond to the recurrence coef\/f\/icients of orthogonal polynomials that are naturally associated to these systems. The analytic models in fact correspond to orthogonal polynomials that have been fully characterized~\cite{5} and they are thus referred to by the name of the polynomials that are attached to them. We have already hinted at that above when we mentioned the Krawtchouk model.

It has more recently been observed that spin chains and waveguide arrays can also exhibit another phenomenon called fractional revival (FR) which is of signif\/icant interest as it relates to entanglement generation. Fractional revival \cite{25,Kay2010} occurs when an initial wavepacket evolves so as to reproduce itself periodically at certain locations~\cite{23,24,de2004quantum,Vinet2015,kay2016tailoring,22}. We shall here be interested in FR at two sites along a spin chain whereby a spin up initially at one end will evolve so as to have, after some FR time, a non-zero probability amplitude only at both ends of the chain. In the photonic picture, this means that the electromagnetic f\/ield amplitude is non-zero only in the f\/irst and last waveguide. Using the qubit correspondence, this means that after propagating a distance $Z_{\rm FR}$ the state $\ket{10\dots 0 }$ has evolved into the entangled state $\alpha \ket{10\dots 0 }+\beta \ket{0\dots01}$.

Interestingly the exploration of the conditions for this FR at two sites has led to the discovery of a new family of orthogonal polynomials \cite{2}, named the para-Krawtchouk polynomials, that have been completely characterized and this has hence provided a model which is the paradigm example of a device with FR at the two ends \cite{1}. The FR properties of other analytic models known to have PST have recently been examined; these results will be succinctly reported here. It would of course be quite interesting to realize FR experimentally in photonic lattices in a way similar to what was done for PST.

When using photonic lattices, the restriction to nearest-neighbour (NN) couplings is clearly an approximation. As a mean to explore the validity of this simplif\/ication and to determine if new phenomenology can arise when the approximation is improved~\cite{kay2006perfect,szameit2007optical}, an analytic model with next-to-nearest couplings and based on the simplest Krawtchouk model has been constructed~\cite{8}. It will be discussed towards the end.

In a nutshell the goal of this short review article is: $1)$~to summarize how orthogonal polynomials intervene in the characterization of photonic lattices with FR and PST; $2)$ to go over a selection of recent results in this respect and $3)$ to of\/fer a compact catalog of analytic models with these properties.

The outline of the paper is as follows. The connection between photon propagation in wave\-guide arrays and orthogonal polynomials is set up in Section~\ref{section2}. The special conditions for FR and PST in these ensembles of waveguides when the NN approximation is made are given in Section~\ref{section3}. How the specif\/ications of the lattices are determined from these conditions using orthogonal polynomials is brief\/ly indicated in Section~\ref{section4}. Section~\ref{section5} which is the bulk of the paper goes over a number of analytic models and discusses their FR and PST properties. Some material in this section is original. Section~\ref{section6} describes one analytic model with next-to-nearest neighbour couplings and comments on extensions and generalizations. A short conclusion completes the paper.

\section{Transport in waveguide arrays and orthogonal polynomials}\label{section2}

Spin chains and photonic lattices are usefully analyzed with the help of orthogonal polynomials (see for instance \cite{efremidis2005revivals,Kay2010,petrovic2015multiport,Rodriguez2011, 3}). Let us consider an array of $(N+1)$ evanescently coupled optical waveguides. Denote by $z$ the propagation distance and by $E_n$ the modal f\/ield amplitude in the $n$\textsuperscript{th} waveguide. In the nearest-neighbour approximation of coupled mode theory \cite{huang1994coupled,rodriguez2015quantum}, the propagation is governed by
 \begin{gather}
 i\frac{{\rm d}}{{\rm d}z}E_n(z) =J_{n+1} E_{n+1}(z) + B_n E_n(z)+J_{n} E_{n-1}(z),\nonumber\\
 n = 0,1,\dots,N, \qquad J_0 = J_{N+1}=0, \label{ham}
 \end{gather}
 $B_n$ is the propagation constant in the $n$\textsuperscript{th} waveguide and $J_{n+1}$, a positive real number, is the evanescent coupling strength between the waveguides $n$ and $n+1$. This $J_{n+1}$ is (approximately) given by
 \begin{gather} \label{coupling}
 J_{n+1} = Ae^{-C d_{n,n+1}},
 \end{gather}
 where $d_{n,n+1}$ is the distance between the $n$\textsuperscript{th} and ($n+1$)\textsuperscript{th} waveguides and $A$, $C$ are measurable quantities that specify the array. Upon introducing the natural basis of $\mathbb{C}^{N+1}$ made out of the vectors
 \begin{gather*}
 \ket{n} = (0,\dots ,0,1,0,\dots ,0)^T, \qquad n =0,\dots,N
 \end{gather*}
 with 1 in the $n$\textsuperscript{th} position and positing
 \begin{gather*}
 \ket{E} = \sum_{n=0}^NE_n(z)\ket{n},
 \end{gather*}
 one can rewrite (\ref{ham}) in the form
 \begin{gather*}
 i\frac{{\rm d}}{{\rm d}z}\ket{E}=J\ket{E},
 \end{gather*}
 where $J$ is the operator def\/ined by
 \begin{gather*}
 J\ket{n} = J_{n+1}\ket{n+1}+B_n\ket{n}+J_n\ket{n-1},
 \end{gather*}
 again with $J_0=J_{N+1} = 0$. It thus follows that
 \begin{gather*}
 \ket{E(z)} = e^{-izJ}\ket{E(0)},
 \end{gather*}
 where in the basis $\{\ket{n}\}$, $J$ is the tridiagonal matrix
 \begin{gather*}
 J=
 \begin{pmatrix}
 B_0	&	J_1	&	0	& 	&
 \\
 J_1	&	B_1	&	J_2	&	&
 \\
 	 0 &	J_2	&	B_2	& \ddots &
 \\
 	 & 		& \ddots		& 	\ddots & J_N
 \\
 & & &J_{N}&B_{N}
 \end{pmatrix}.
 \end{gather*}
 The Jacobi matrix $J$ is Hermitian and has non-degenerate eigenvalues $\lambda_s$, $s = 0,1,\dots,N$ when its entries are non-negative. Let us denote the eigenvectors of $J$ by $\ket{\lambda_s}$:
 \begin{gather*}
 J\ket{\lambda_s} = \lambda_s\ket{\lambda_s}.
 \end{gather*}
 We shall now indicate how orthogonal polynomials are naturally associated to the dynamics. This simply follows from the fact that Jacobi matrices are diagonalized by orthogonal polynomials. Consider the expansion of the eigenbasis $\{\ket{\lambda_s}\}$ in term of the natural basis $\{\ket{n}\}$:
 \begin{gather*} 
 \ket{\lambda_s} = \sum_{n=0}^N W_{s,n} \ket{n}.
 \end{gather*}
 Acting on both sides with $J$, we get
 \begin{gather*}
 \lambda_s\sum_{n=0}^N W_{s,n}\ket{n} = \sum_{n=0}^NW_{s,n}(J_{n+1}\ket{n+1}+B_n\ket{n}+J_n\ket{n-1})
 \end{gather*}
 and since the natural basis is orthonormal, we have
 \begin{gather*}
 \lambda_sW_{s,n} = J_{n+1}W_{s,n+1}+B_nW_{s,n}+J_nW_{s,n-1}.
 \end{gather*}
 Taking $W_{s,n}$ in the form
 \begin{gather*} 
 W_{s,n} =W_{s0}\chi_n(\lambda_s)
 \end{gather*}
 we see that the coef\/f\/icients $\chi_n(x)$, $n=0,1,\dots,N$, constitute an orthogonal polynomial set owing to the fact that they satisfy the three term recurrence relation
 \begin{gather*}
 x\chi_n(x) = J_{n+1}\chi_{n+1}(x)+B_n\chi_n(x)+ J_{n}\chi_{n-1}(x)
 \end{gather*}
 with the initial condition $\chi_{-1}(x) = 0$. Since both bases are orthonormal, $\braket{s}{s'} = \delta_{ss'}$ and $\braket{n}{m} = \delta_{nm}$ it follows that the matrix ($W_{s,n}$) is orthogonal
 \begin{gather*}
 \sum_{n=0}^N W_{s,n}W_{s',n} = \delta_{ss'}, \qquad\sum_{s=0}^N W_{s,n}W_{s,m} = \delta_{nm}.
 \end{gather*}
 The reverse expansion is thus given by
 \begin{gather} \label{nas}
 \ket{n} = \sum_{s=0}^N W_{s0} \chi_n(\lambda_s)\ket{\lambda_s}
 \end{gather}
 and one explicitly recovers from $\braket{n}{m} = \delta_{nm}$ that the functions $\chi_n(\lambda_s)$ are normalized orthogonal polynomials in the discrete variable $\lambda_s$. The orthogonality relation reads
 \begin{gather*} 
 \sum_{s=0}^Nw_s\chi_n(\lambda_s)\chi_m(\lambda_s) = \delta_{nm}
 \end{gather*}
 with $w_s = W_{s0}^2$ playing the role of the weight.

We can def\/ine the monic polynomials $P_n(x)$
 \begin{gather*}
P_n(x) = J_1\cdots J_n\chi_n(x).
 \end{gather*}
 Furthermore the characteristic polynomial $P_{N+1}(x)$
 \begin{gather*}
 P_{N+1}(x) = (x-\lambda_0)(x-\lambda_1)\cdots (x-\lambda_N)
 \end{gather*}
 is manifestly orthogonal to the polynomials $\chi_n(x)$, $n = 0,\dots,N$. From the general theory of orthogonal polynomials, it is known that \cite{chihara} the discrete weights $w_s$ can be written in the form
 \begin{gather} \label{19}
 w_s = \frac{\sqrt{h_N}}{\chi_N(\lambda_s)P'_{N+1}(\lambda_s)}, \qquad s=0,1,\dots,N,
 \end{gather}
 where
 \begin{gather*}
 \sqrt{h_N} = J_1J_2\cdots J_{N}
 \end{gather*}
 and where $P'_{N+1}(x)$ stands for the derivative of $P_{N+1}(x)$. If one takes the eigenvalues in increasing order,
 \begin{gather*}
 \lambda_0<\lambda_1<\cdots<\lambda_N,
 \end{gather*}
 it is easy to see that
 \begin{gather} \label{18}
 P'_{N+1}(\lambda_s) = (-1)^{N+s}|P'_{N+1}(\lambda_s)|.
 \end{gather}

\section[Fractional revival (FR) and perfect state transfer (PST) in the nearest-neighbour approximation]{Fractional revival (FR) and perfect state transfer (PST)\\ in the nearest-neighbour approximation}\label{section3}

 To achieve PST and FR, some conditions must be respected. These, in fact, will only depend on the spectrum of $J$. PST will be achieved if a photon initially in the waveguide $0$ will be found in the waveguide $N$ after a propagation distance $Z_{\rm PST}$, in other words PST will be observed if
 \begin{gather} \label{evo}
 e^{-iZ_{\rm PST}J}\ket{0}=e^{i\phi}\ket{N}
 \end{gather}
 with $\phi$ an arbitrary phase.

 Using the expansion (\ref{nas}) of the natural basis into the eigenbasis and remembering that $\chi_0(x)=1$, one sees that (\ref{evo}) implies
 \begin{gather} \label{cond}
 e^{-iZ_{\rm PST}\lambda_s}=e^{i\phi} \chi_N(\lambda_s).
 \end{gather}
 This shows that $\chi_N(\lambda_s)$ is a phase, but since it is real we must have
 \begin{gather*}
 \chi_N(\lambda_s) = \pm 1.
 \end{gather*}
 Owing to the fact that the zeros of $\chi_N(x)$ must interlace the zeros, $\lambda_s$, of $P_{N+1}(x)$, the sign of~$\chi_N(\lambda_s)$ must alternate, i.e., $\chi_N(\lambda_s) \propto(-1)^s$. Now since $w_s$ as given by (\ref{19}) must be positive, in view of (\ref{18}), we see that
 \begin{gather} \label{al}
 \chi_N(\lambda_s) = (-1)^{N+s}.
 \end{gather}
 As shown in \cite{Kay2010}, this condition is tantamount to the matrix $J$ being symmetric with respect to the anti-diagonal, i.e., to the relations
 \begin{gather} \label{bj}
 B_{N-n} = B_n,\qquad J_{N-n+1} = J_n.
 \end{gather}
 The condition (\ref{al}) or equivalently (\ref{bj}) is thus necessary for the occurrence of PST.

 Making use of this necessary condition in (\ref{cond}) one f\/inds that PST will take place if
 \begin{gather*}
 e^{-i(Z_{\rm PST}\lambda_s+\phi)} = e^{i\pi(N+s)}.
 \end{gather*}
 This will be satisf\/ied if
 \begin{gather} \label{condspec}
 Z_{\rm PST}\lambda_s+\phi+\pi(N+s) = 2\pi L_s,
 \end{gather}
 where $L_s$ is a sequence of integers that may depend on $s$.

 It follows from (\ref{condspec}) that the dif\/ference between 2 successive eigenvalues is given by
 \begin{gather} \label{condPST}
 \lambda_s-\lambda_{s-1}=\frac{\pi}{Z_{\rm PST}}M_s,
 \end{gather}
 where $M_s$ is an odd integer.

 Together (\ref{bj}) and (\ref{condPST}) are necessary and suf\/f\/icient for PST. As announced, these are conditions on the spectrum of $J$.

 Let us now turn to FR in the two external waveguides. This will occur if there is a distance~$Z_{\rm FR}$ at which the only non-zero amplitudes are $E_0(Z_{\rm FR})$ and $E_N(Z_{\rm FR})$. In this case we shall have
 \begin{gather} \label{fr}
 e^{-iZ_{\rm FR}J}\ket{0}=\mu \ket{0}+\nu \ket{N}
 \end{gather}
 with the normalization condition
 \begin{gather*}
 |\mu|^2+|\nu|^2=1.
 \end{gather*}
 When this is so, the wavepacket initially localized in the waveguide $0$ is revived in the wave\-gui\-de~$0$ and~$N$. In the special case
 \begin{gather*}
 \mu = \nu = \frac{1}{\sqrt{2}}
 \end{gather*}
 we have an analog of the maximally entangled state $\ket{E} = \frac{E_0(0)}{\sqrt{2}} (\ket{10\dots0}+\ket{00\dots1} )$ where the vectors $\ket{1}$, $\ket{0}$ denote the presence ($\ket{1}$) or the absence ($\ket{0}$) of a photon in the waveguide.
 One also sees that PST is a special case of FR with
 \begin{gather*}
 \mu = 0,\qquad \nu = e^{i\phi}.
 \end{gather*}
 As for PST, the relation (\ref{fr}) can be translated into a spectral condition by using the expansion~(\ref{nas}). This leads to
 \begin{gather} \label{fr2}
 e^{-iZ_{\rm FR}\lambda_s} = \mu+\chi_N(\lambda_s)\nu.
 \end{gather}
In the following we shall look for the occurrence of FR in systems with mirror symmetry~(\ref{bj}). We shall thus assume that the necessary condition~(\ref{al}) for PST is satisf\/ied. As a result~(\ref{fr2}) will read
 \begin{gather} \label{fr3}
 e^{-iZ_{\rm FR}\lambda_s} = e^{i\phi}(\mu'+(-1)^{N+s}\nu'),
 \end{gather}
 where we have factored out the phase $e^{i\phi}$ of $\mu$ implying that $\mu = e^{i\phi}\mu'$ with $\mu'$ real and $\nu = e^{i\phi}\nu'$. For consistency, the norm of ($\mu'+(-1)^{N+s}\nu'$) must be $1$ and with $\mu'$ real this requires $\nu'$ to be purely imaginary. Since $|\mu|^2+|\nu|^2 = 1$, we may thus take the following parametrization
 \begin{gather*}
 \mu' = \cos\theta, \qquad \nu' = i\sin\theta.
 \end{gather*}
 Observe then that PST ($\mu =0$) corresponds to $\theta = \frac{\pi}{2}$. As in the case for PST, condition (\ref{fr3}) can be seen to imply restrictions on successive eigenvalues:
 \begin{gather} \label{cond2}
 Z_{\rm FR}(\lambda_{2s}-\lambda_{2s-1})=-(-1)^N2\theta+2\pi L_s^{(0)},\\
 \label{Cond2*} Z_{\rm FR}(\lambda_{2s+1}-\lambda_{2s})=(-1)^N2\theta+2\pi L_s^{(1)},
 \end{gather}
 where $L^{(i)}_s$ are arbitrary sequences of integers that depend on $s$.

 FR will occur if those conditions are respected for all eigenvalues of the spectrum. Note that, as shown in \cite{1}, in this mirror-symmetric situation, FR also occurs between symmetrically positioned sites say between~$\ket{n}$ and~$\ket{N-n}$.
\section{Engineering waveguide arrays with FR and PST}\label{section4}
The engineering of photonic lattices with FR and PST involves an inverse spectral problem, as f\/irst discussed in~\cite{karbach2005spin}. Given a set of $\{\lambda_s\}$ that satisfy the FR/PST conditions, we must f\/ind the mirror-symmetric matrix~$J$ formed out of the coupling and propagation constants that has the $\{\lambda_s\}$ for eigenvalues. This is done by constructing the orthogonal polynomials associated to the problem whose recurrence coef\/f\/icients will provide the characteristics of the array that we are looking for. Once these have been obtained, it is a matter of creating appropriately the waveguides and of choosing the distances between them to make the specif\/ications of the lattice match those prescribed theoretically.

 Let us brief\/ly recall how this inverse spectral problem is solved from the construction of the associated orthogonal polynomials. First, observe that the condition $\chi_N(\lambda_s)=(-1)^{N+s}$ f\/ixes the values of a polynomial of degree $N$ at $N+1$ points and thereby completely determines $\chi_N(x)$. This polynomial can be explicitly constructed using the Lagrange interpolation polynomials:
 \begin{gather*}
 \chi_N(x) = \sum_{s=0}^N (-1)^{N+s}\mathcal{L}_s.
 \end{gather*}
 Here $\mathcal{L}_s$ is the standard Lagrange polynomial
 \begin{gather*}
 \mathcal{L}_s = \prod_{i = 0}^{N'}\frac{x-\lambda_i}{\lambda_s-\lambda_i},
 \end{gather*}
 where the $'$ stands for $i\neq s$.

The monic polynomial $P_N(x)$ is then obtained by dividing $\chi_N(x)$ by the coef\/f\/icient of $x^N$. The spectral data $\{\lambda_s\}$ provides another member of the set of orthogonal polynomials, namely, the characteristic polynomial $P_{N+1}(x) = (x-\lambda_0)\cdots(x-\lambda_N)$. The recurrence relation for the monic polynomials $P_n(x) = J_1\cdots J_n\chi_n(x)$ is
 \begin{gather*}
 xP_n(x) = P_{n+1} + B_nP_n(x) +U_nP_{n-1}(x)
 \end{gather*}
with $U_n = J_n^2$. For all these polynomials $P_n(x)$, the coef\/f\/icients of the leading monomial~$x^n$ is~$1$.

We thus have
 \begin{gather} \label{norm}
 P_{N+1}(x) = (x-B_N)P_N(x)-J_N^2P_{N-1}.
\end{gather}
 We know $P_{N+1}$ and $P_N$. If we look at (\ref{norm}) degree by degree, we see that the terms in $x^{N+1}$ are equal. Then the coef\/f\/icient of~$x^N$ in $P_{N+1}$ gives $-B_N$. Now the coef\/f\/icients of $x^{N-1}$ in $P_{N+1}(x)-(x-B_N)P_N(x)$ gives $-J_N^2$ and at this point we know~$B_N$,~$J_N^2$ and~$P_{N-1}(x)$. We can then iterate to f\/ind $B_{N-1}$, $J_{N-1}^2$ and $P_{N-2}(x)$ and so on. In the end the mirror-symmetric matrix~$J$ is fully characterized and hence provides all the couplings~$J_n$ and propagation cons\-tants~$B_n$ for $n=0,\dots,N$.

While the algorithm described above is conceptually straightforward it might not be the most ef\/f\/icient to obtain the parameters numerically. Another method is given in Gladwell~\cite{Gladwell2004}; the approach described in~\cite{genest2017persymmetric} based on persymmetric polynomials is also generally faster.

\section{A review of analytic models with of FR and PST}\label{section5}
 In a number of cases, spectra that satisfy (\ref{condPST}) and (\ref{cond2}) have been found to correspond to orthogonal polynomials that can be fully characterized and this has therefore led to models that can be described analytically. These arrays are often referred to by the name of the orthogonal polynomials associated to them and turn out to be most useful. They have f\/irst been identif\/ied for their PST properties. In the following we shall go over many of these systems. We shall give in each case the corresponding spectrum $\{\lambda_s\}$ and provide the couplings $J_n$ and propagation constants $B_n$ that the algorithm described in Section~\ref{section4} gives. We shall indicate what the FR and PST conditions entail. For FR we shall provide the mixing angles $\theta$ that are possible as well as the distances $Z_{\rm FR}$ at which this FR occurs. We shall also record the restrictions that must be imposed for PST to happen also. To allow for an adjustment of the reference strength of the couplings we shall introduce a parameter~$\beta$ as a global factor in the spectra: $\lambda_s\rightarrow\beta\lambda_s$.
 \subsection{Krawtchouk}

The Krawtchouk model is the simplest analytic model that has been designed \cite{12}. For a f\/ixed maximum coupling strength, it exhibits shortest state transfer time and thus minimizes \cite{yung2006quantum} the risk of introducing noise. This in part explains why it has been favored in experimental validations \cite{9,10,18}. As we shall see this system does not admit FR however. The Krawtchouk array emerges when the linear spectrum
 \begin{gather} \label{kra}
 \lambda_s =\beta\left( s-\frac{N}{2}\right)
 \end{gather}
 is considered.

 The couplings between the waveguides take the form
 \begin{gather} \label{jkraf}
 J_n = \frac{\beta}{2}\sqrt{n(N-n+1)}
 \end{gather}
 and the propagation constant in the $n$\textsuperscript{th} waveguide is
 \begin{gather*}
 B_n = 0.
 \end{gather*}
 Of course the propagation constants can be shifted by a uniform value by adding a global term to the spectrum.
 \subsubsection{FR}
 For the spectrum (\ref{kra}), the sequences of integers do not depend on $s$, $L_s^{(i)}=c_i$ and the restrictions~(\ref{cond2}) amount to
 \begin{gather*}
 Z_{\rm FR}\beta = -(-1)^N2\theta+2\pi c_0,\\
 Z_{\rm FR}\beta = (-1)^N2\theta+2\pi c_1,
 \end{gather*}
 where $c_0$ and $c_1$ are integers. Solving for $\theta$ and $Z_{\rm FR}$ yields
 \begin{gather*}
 \theta = (-1)^N\frac{\pi}{2}(c_0-c_1)
 \end{gather*}
 and
 \begin{gather*}
 Z_{\rm FR} = \frac{\pi}{\beta}(c_0+c_1).
 \end{gather*}
 We thus observe that FR does not occur. Indeed $\theta$ must be a multiple of $\frac{\pi}{2}$ and as already observed this can only lead to PST or Perfect Return. Moreover, since $(c_0-c_1)$ has the same parity as $(c_0+c_1)$, PST will happen for $Z_{\rm PST} = \frac{\pi q}{\beta}$ where~$q$ is an odd integer and Perfect Return will occur when~$q$ is even.

 \subsection{para-Krawtchouk}
 The para-Krawtchouk array is the prototype of systems with FR \cite{1}. It involves a parame\-ter~$\delta$ that determines the mixing angle~$\theta$. The associated para-Krawtchouk polynomials are not conventional polynomials; they were discovered as a result of investigations of systems with FR and PST~\cite{2}. It is remarkable that a full characterization of the polynomials could be carried out.

 The spectrum here is
 \begin{gather*}
 \lambda_s = \beta\left(s-\frac{N-1+\delta}{2}+\frac{(\delta-1)}{2}\big(1-(-1)^s\big)\right), \qquad s=0,1,\dots,N,
 \end{gather*}
 where $0<\delta<2$ is a real parameter. It is referred to as a bi-lattice spectrum as one observes that it is the superposition of two linear lattices shifted by $\delta$. One notes also that the linear lattice~(\ref{kra}) is recovered when $\delta =1$.

 The mirror symmetric couplings and propagation constants that are obtained by applying the Euclidian algorithm of Section~\ref{section4} to this spectrum are the following
 \begin{gather*}
 J_n = \frac{\beta}{2}\sqrt{\frac{n(N+1-n)[(N+1-2n)^2-\delta^2]}{(N-2n)(N-2n+2)}},\qquad B_n = 0
 \end{gather*}
 for $N$ odd, and
 \begin{gather*}
 J_n = \frac{\beta}{2}\sqrt{\frac{n(N+1-n)[(2n-N-1)^2-(\delta-1)^2]}{(2n-N-1)^2}},\\
 B_n = \frac{(\delta-1)(N+1)}{4}\left(\frac{1}{2n-N-1}-\frac{1}{2n+1-N}\right)
 \end{gather*}
 for $N$ even.

 Note that the propagation constants must vary from waveguide to waveguide when $N$ is even. It is seen that $J_n$ and $B_n$ verify the mirror symmetry property (\ref{bj}) and that, as should be, the Krawtchouk coupling are retrieved when $\delta=1$.
 \subsubsection{FR}
 When specialized to the para-Krawtchouk spectrum, the sequences of integers do not depend on $s$, $L_s^{(i)}=c_i$ and the conditions (\ref{cond2}) and~(\ref{Cond2*}) become
 \begin{gather} \label{cpk1}
 Z_{\rm FR}\beta(2-\delta)=-2(-1)^N\theta+2\pi c_0,\\
 \label{cpk2} Z_{\rm FR}\beta\delta=2(-1)^N\theta+2\pi c_1,
 \end{gather}
 where $c_0$ and $c_1$ are integers.

 Expressions for $\theta$ and $Z_{\rm FR}$ can be obtained from (\ref{cpk1}) and (\ref{cpk2}). One f\/inds
 \begin{gather*}
 \theta=(-1)^N\left[-\pi c_1+\frac{\pi\delta}{2}(c_0+c_1)\right]
 \end{gather*}
 showing that $\delta$ directly determines $\theta$ and for $Z_{\rm FR}$ one gets
 \begin{gather} \label{zfrpk}
 Z_{\rm FR}=\frac{\pi}{\beta}(c_0+c_1).
 \end{gather}
 These results indicate that FR happens in this model and that the mixing angle $\theta$ is prescribed by the value of $\delta$. The distances $Z_{\rm FR}$ are integer multiples of $\frac{\pi}{\beta}$.

 \subsubsection{PST}
 To investigate whether the para-Krawtchouk model admits PST in addition to FR, one examines the relation that the PST condition (\ref{condPST}) further imposes. One f\/inds here
 \begin{gather} \label{cond2pk}
 \beta(1-(\delta-1)(-1)^s)=\frac{\pi}{Z_{\rm PST}}M_{\epsilon},\qquad \epsilon=(-1)^s.
 \end{gather}
The sequences $M_{\epsilon}$ of odd integers here only depend on the parity $\epsilon$ of $s$. We distinguish these two parities and write $M_{+1}=2c_0+1$ and $M_{-1}=2c_1+1$ with $c_0$ and $c_1$ integers. Equa\-tion~(\ref{cond2pk}) then translates into
 \begin{gather} \label{zpstpk}
 Z_{\rm PST}=\frac{\pi}{\beta}\frac{2c_0+1}{2-\delta}
 \end{gather}
 for $s$ even, and
 \begin{gather} \label{zpstpk2}
 Z_{\rm PST}=\frac{\pi}{\beta}\frac{2c_1+1}{\delta}
 \end{gather}
 for $s$ odd.

 In order for the right hand sides of (\ref{zpstpk}) and (\ref{zpstpk2}) to be equal one must have
 \begin{gather} \label{condelta}
 \delta = \frac{(2c_1+1)}{(c_0+c_1+1)}.
 \end{gather}
 This says that for PST to occur in addition to FR in the para-Krawtchouk model, the parameter~$\delta$ must be a rational number of the form~$\delta =\frac{p}{q}$ where $p$ and $q$ are coprime and $p$ is odd.

 The minimal distance for PST is obtained when $c_1=0$ in which case $Z_{\rm PST}=\frac{\pi}{\beta\delta}$. In general the distance for PST is
 \begin{gather*}
 Z_{\rm PST}=\frac{\pi}{\beta}(c_0+c_1+1).
 \end{gather*}
 This should be compared with the expression for $Z_{\rm FR}$ given in~(\ref{zfrpk}): provided $\delta$ satisf\/ies~(\ref{condelta}), if FR occurs at $Z_{\rm FR}=\frac{\pi}{\beta}$, PST will happen at $Z_{\rm PST}=2Z_{\rm FR}$.

 \subsection{dual-Hahn}
 The dual-Hahn system has been identif\/ied early on \cite{4}. It is based on the quadratic spetrum
 \begin{gather} \label{sphahn}
 \lambda_s = \beta s(s+2\gamma+1)
 \end{gather}
 with a parameter $\gamma>-1$.

 Its coupling and propagation constants are given by
 \begin{gather*}
 J_n = \beta\sqrt{(n+1)(N-n)(\gamma+N-n)(\gamma+n+1)},\\
 B_n = 2n(N-n) + (\gamma+1)N.
 \end{gather*}
 \subsubsection{FR}
 For $s$ even, the FR conditions (\ref{cond2}) and (\ref{Cond2*}) require $L_s^{(0)}=c_1 s+c_0$ and yield
 \begin{gather*}
 (2s+\gamma)\beta Z_{\rm FR} = -(-1)^N\theta+\pi(c_1 s+c_0),
 \end{gather*}
 where $c_1$ and $c_0$ are an integers. This implies
 \begin{align} \label{zfrhp}
 (i)\quad & \beta Z_{\rm FR} = \frac{\pi}{2} c_1,\\
 \label{zfrhp2} (ii)\quad &\gamma \beta Z_{\rm FR} = \pi c_0-(-1)^N\theta.
 \end{align}
 For $s$ odd, with $L_s^{(1)}=c_1' s+c_0'$ one obtains
 \begin{gather*}
 (2s+1+\gamma)\beta Z_{\rm FR} = (-1)^N\theta+\pi(c_1' s+c_0'),
 \end{gather*}
 where $c_1'$ and $c_0'$ are both integers. One has then
 \begin{align} \label{zfrhi}
 (i)\quad&\beta Z_{\rm FR} = \frac{\pi}{2} c_1', \\
 \label{zfrhi2} (ii)\quad&\beta(1+\gamma) Z_{\rm FR} = (-1)^N\theta+\pi c_0'.
 \end{align}
 The compatibility of the equations (\ref{zfrhp}), (\ref{zfrhp2}) and (\ref{zfrhi}), (\ref{zfrhi2}) leads to
 \begin{gather*}
 \theta=(-1)^N \left(\frac{\pi c_0(1+\gamma)-\gamma\pi c_0'}{2\gamma+1}\right).
 \end{gather*}
 One further f\/inds
 \begin{gather*}
 \gamma=\frac{2(c_0+c_0')-c_1}{2c_1}
 \end{gather*}
 and
 \begin{gather*}
 \theta = (-1)^N\left(\frac{\pi c_1}{4}+\frac{\pi(c_0-c_0')}{2}\right).
 \end{gather*}
 The main features of this model are thus the following. Balanced FR may occur. By balanced FR we mean a situation where amplitudes in the f\/irst and last waveguides are equal in magnitude, this correspond to~$\theta$ being equal to odd multiples of~$\frac{\pi}{4}$. We indeed see that this is possible when $c_1=q$ is odd. The FR distance $Z_{\rm FR}$ is given by
 \begin{gather*}
 Z_{\rm FR}=\frac{\pi}{2\beta}q.
 \end{gather*}
 As we shall indicate below the dual-Hahn system is a special case of the para-Racah model that also admits FR.
 \subsubsection{PST}
 When $\lambda_s$ is given by (\ref{sphahn}) the PST condition (\ref{condPST}) requires $M_s=2(c_1s+c_0)+1$ and becomes
 \begin{gather} \label{cohahnpst}
 \beta (s +\gamma) = \frac{\pi}{2Z_{\rm PST}}(2(c_1 s+c_0 )+1)
 \end{gather}
 where $c_1$ and $c_0$ are integers. Equating each power of~$s$ in~(\ref{cohahnpst}) yields
 \begin{gather*}
 Z_{\rm PST} = \frac{\pi}{\beta}c_1, \qquad
 Z_{\rm PST} = \frac{\pi}{2\beta \gamma}(2c_0 +1).
 \end{gather*}
 The consistency of these two relations requires that
 \begin{gather*}
 \gamma=\frac{2c_0+1}{2c_1}.
 \end{gather*}
 The minimal distance for PST is $Z_{\rm PST}=\frac{\pi}{\beta}$ with $\gamma = \frac{1}{2}$ ($c_0=0$, $c_1=1$).

PST will thus occur if the parameter $\gamma$ is of the form $\frac{p}{2q}$ with~$p$ and~$q$ coprime integers and~$p$ odd. Note that then $Z_{\rm PST} = 2Z_{\rm FR}$.

 \subsection[Special $q$-Racah]{Special $\boldsymbol{q}$-Racah}
We now consider an array associated to a special case of $q$-Racah polynomials \cite{5}. Consider the set of eigenvalues given by the following exponential lattice:
 \begin{gather} \label{qrsp}
 \lambda_s = \beta\big(q^{-s+N/2}-q^{s-N/2}\big), \qquad s=0,\dots,N,
 \end{gather}
 where $q$ is a real number between $0$ and $1$. This corresponds to a $q$-deformation of the Krawtchouk spectrum since $\lim\limits_{q\to 1^-} \frac{-\lambda_s}{q-q^{-1}}=\beta(s-\frac{N}{2})$. Some restrictions will be added to allow for FR and PST~\cite{3}. It is observed that the $\lambda_s$ def\/ined by~(\ref{qrsp}) obey the three-term recurrence relation
 \begin{gather} \label{rq}
 \lambda_{s} = \big(q+q^{-1}\big)\lambda_{s-1}-\lambda_{s-2}.
 \end{gather}
 Take
 \begin{gather} \label{k}
 q+q^{-1} = K
 \end{gather}
 with $K$ an integer. We shall assume that $K$ is greater than $2$ because for $K=2$ we have $q=1$ and we then recover the Krawtchouk spectrum. Relation~(\ref{k}) implies that
 \begin{gather*}
 q = \frac{K}{2}-\sqrt{\frac{K^2}{4}-1}.
 \end{gather*}
 The coupling and propagation constants are here given by
 \begin{gather*}
 J_n = \beta\sqrt{\frac{\big(1-q^{2n}\big)\big(q^{2(n-N-1)}-1\big)}{\big(1+q^{2n-N-2}\big)\big(1+q^{2n-N}\big)}}, \qquad B_n = 0.
 \end{gather*}
 \subsubsection{FR}
 With the help of (\ref{rq}), the spectral conditions (\ref{cond2}) and (\ref{Cond2*}) for FR can be rewritten as follows
 \begin{gather} \label{fr9}
 Z_{\rm FR}(K-2)\lambda_{2s} = (-1)^N4\theta+2\pi L_s^{(0)},\qquad
 Z_{\rm FR}(K-2)\lambda_{2s+1} = -(-1)^N4\theta+2\pi L_s^{(1)}.\!\!\!
 \end{gather}
 For $N$ even, observe that $\lambda_{\frac{N}{2}}=0$. This leads to
 \begin{gather*}
 \theta = \frac{\pi }{2}L_{\frac{N}{2}}.
 \end{gather*}
 Hence, for $N$ even, only PST can be achieved.

 For $N$ odd, no restrictions of this type happen. Consider here an alternative way of wri\-ting~(\ref{fr3}):
 \begin{gather} \label{fr10}
 Z_{\rm FR} \lambda_{2s} = -\phi -(-1)^N\theta + 2\pi L_s^{(2)}, \qquad
 Z_{\rm FR} \lambda_{2s+1} = -\phi +(-1)^N\theta + 2\pi L_s^{(3)}.
 \end{gather}
 From (\ref{fr9}) and (\ref{fr10}) one obtains for the mixing angle
 \begin{gather} \label{theta}
 \theta = (-1)^N\left(\frac{2\pi Q_s}{K+2}-\phi\frac{K-2}{K+2}\right),
 \end{gather}
 where $Q_s$ is a sequence of integers that depend on~$s$. This indicates that FR can happen in the special $q$-Racah array when~$N$ is odd for a variety of mixing angle $\theta$ related to~$K$. This is a new observation as far as we know.
 \subsubsection{PST}
 In order to enforce the PST condition, one requires that all $\lambda_s$ are integers with alternating parity. To achieve PST, $K$ must be an even integer if~$N$ is even and $K = 6,10,14,\dots$ if~$N$ is odd. This is in keeping with~(\ref{theta}).
\begin{figure}[h!] \centering
 \includegraphics[width=0.8\linewidth]{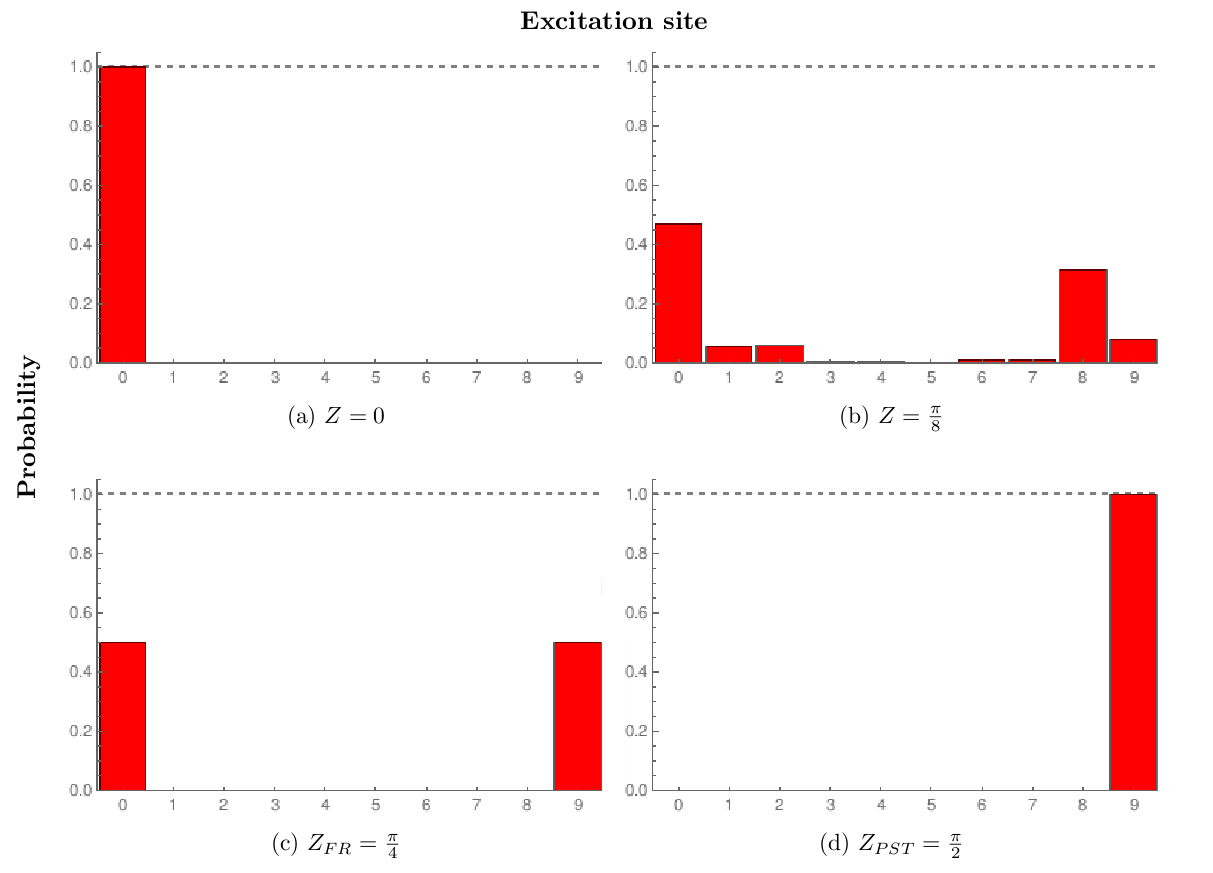}
 \caption{Probability of f\/inding a single photon at each site of a $10$-waveguide array ($N=9$) of $q$-Racah type with parameters $\beta=(K-2)^{-1/2}$ and $K=6$ for various distances. (a)~The photon is initially inserted at site $0$, (b)~after some distance $Z=\frac{\pi}{8}$ the photon spreads out through the lattice and (c)~fractional revival is observed at $Z_{\rm FR}=\frac{\pi}{4}$ and (d)~perfect state transfer observed at distance $Z_{\rm PST}=\frac{\pi}{2}= 2Z_{\rm FR}$.} \label{figb:dd}
\end{figure}

 \subsection{para-Racah}
 A novel family of orthogonal polynomials corresponding to an alternative truncation of the Wilson polynomials has been identif\/ied recently \cite{32}. It is associated to the quadratic bi-lattice def\/ined by
 \begin{gather*}
 \lambda_{2s} = \beta(s+a)^2,\qquad \lambda_{2s+1} = \beta(s+c)^2, \qquad s=0,\dots,N,
 \end{gather*}
 where $a >-\frac{1}{2}$, $|a|<c<|a+1|$ and both $c$ and $a$ are real.

Note that the set $\{\lambda_s\}$ can be viewed as the superposition of two quadratic lattices shifted by $c-a$. One observes also that an equivalent quadratic lattice of the form~(\ref{sphahn}) is recovered if $c=a+\frac{1}{2}$.

 The recurrence coef\/f\/icients provide the following coupling and propagation constants:
 \begin{gather*} 
 B_n=\frac{1}{2}\left[a(a+j)+c(c+j)+n(N-n)\right], \\
 J_n=\left[\frac{n(N+1-n)(N-n+a+c)(n-1+a+c)(n-j-1)^2-(a-c)^2}{4(N-2n)(N-2n+2)}\right]^{\frac{1}{2}}
 \end{gather*}
 for $N$ odd, and
 \begin{gather*}
 J_n=\left[\frac{n(N+1-n)(n-1+a+c)(N-n+a+c)(n-j+a-c)(n-j+c-a-1)}{4(N-2n+1)^2}\right]^\frac{1}{2},\\
 B_n= \frac{1}{2}\big(a^2+c^2+n-n^2\big)+\frac{1}{4}(2n+a+c)(N-1)+\frac{(n+1)(n+a+c)(1+2a-2c)}{4(1+2n-N)}\\ \hphantom{B_n=}{}+\frac{n(n-1+a+c)(1+2a-2c)}{4(1-2n+N)}
 \end{gather*}
 for $N$ even.

 The array with these specif\/ications has been shown to enact FR and PST \cite{17}.
 \subsubsection{FR}
 The conditions (\ref{cond2}) and (\ref{Cond2*}) when specialized to the para-Racah spectrum require $L_s^{(0)}=c_1s+c_0$ and $L_s^{(1)}=c_1's+c_0'$ and read
 \begin{gather} \label{frpr}
 \beta Z_{\rm FR}(1+a-c)(2s-1+a+c)=-(-1)^N2\theta+2\pi(c_1 s+c_0),\\
 \label{frpr2} \beta Z_{\rm FR}(c-a)(2s+c+a)=(-1)^N2\theta+2\pi(c_1' s+c_0')
 \end{gather}
 where $c_1$, $c_0$, $c_1'$ and $c_0'$ are integers.

 Relations for $Z_{\rm FR}$, $\theta$, $a$ and $c$ are obtained from (\ref{frpr}) and (\ref{frpr2}). First one f\/inds
 \begin{gather*}
 a = \frac{2(c_0+c_0')+(c_1-c_1')}{2(c_1+c_1')}, \qquad
 c = \frac{2(c_0+c_0')+(c_1+c_1')}{2(c_1+c_1')}
 \end{gather*}
 showing that $a$ and $c$ must be rational numbers.

 The formula for $\theta$ is
 \begin{gather*}
 \theta = (-1)^{N+1}\frac{\pi}{2}\left(\frac{2(c_1c_0'-c_0c_1')-c_1c_1'}{c_1+c_1'}\right)
 \end{gather*}
 and the distances are given by
 \begin{gather*}
 Z_{\rm FR}=\frac{\pi}{\beta}(c_1+c_1').
 \end{gather*}
 The para-Racah model thus exhibit FR when $a$ and $c$ are rational. If one wants to retrieve the mixing angle $\theta_{\text{d-H}}$ for the dual-Hahn model, it suf\/f\/ices to take the equivalent condition for $c= a+\frac{1}{2}$ which is $c_1=c_1'$. One gets
 \begin{gather*}
 \theta_{\text{d-H}}= (-1)^N\left(\frac{\pi}{4}c_1+\frac{\pi}{2}(c_0-c_0')\right),
 \end{gather*}
 which is exactly the mixing angle of the dual-Hahn model.
 \subsubsection{PST}
 To investigate if the para-Racah model admits PST in addition to FR, we turn to condi\-tion~(\ref{condPST}) which becomes with $M_s^{(0)}=2(c_1 s+c_0)+1$ and $M_s^{(1)}=2(c_1' s+c_0')+1$
 \begin{gather} \label{pstpr}
 \beta Z_{\rm PST}(1+a-c)(2s+a+c-1)=\pi (2(c_1 s+c_0)+1),\\
 \label{pstpr2} \beta Z_{\rm PST}(c-a)(2s+a+c)=\pi(2(c_1' s+c_0')+1).
 \end{gather}
 Using (\ref{pstpr}) and (\ref{pstpr2}) one obtains
 \begin{gather*}
 a =\frac{1}{2}\left(\frac{2c_0+1}{c_1}+\frac{c_1}{c_1+c_1'}\right), \qquad
 c = \frac{1}{2}\left(\frac{2c_0+1}{c_1}+\frac{c_1'}{c_1+c_1'}+1\right)
 \end{gather*}
 and the PST distance can be written as follows
 \begin{gather*}
 Z_{\rm PST}=\frac{\pi}{\beta}(c_1+c_1').
 \end{gather*}
 This provides the details of the PST occurrences in the para-Racah array.

\section{A model with next-to-nearest neighbour interaction}\label{section6}
In this last section, we wish to indicate how a class of analytic models with next-to-nearest neighbour (NNN) interactions can be constructed from those with NN couplings. The one based on the Krawtchouk recurrence coef\/f\/icients will be presented in some details. As a rule, in settings of the type that have been used experimentally the NN approximation is very good; interestingly however allowing for NNN interactions might show the occurrence of FR even if the restriction of the model to NN couplings does not permit it. Let~$J$ be a tridiagonal mirror-symmetric matrix with spectrum $\lambda_s$, $s=0,\dots,N$, wich is thus such that the necessary condition~(\ref{al}) for PST is satisf\/ied. Let
 \begin{gather*}
 \bar{J} = \alpha J^2 +\beta J.
 \end{gather*}
 It follows that $\bar{J}$ is pentadiagonal. Take the light propagation in the array to be governed by
 \begin{gather*}
 i\frac{{\rm d}}{{\rm d}z}\ket{E}=\bar{J}\ket{E}.
 \end{gather*}
 In terms of amplitudes we thus have the following system of equations with NNN interactions
 \begin{gather}
 i\frac{{\rm d}}{{\rm d}z}E_n = \alpha(J_{n-1}J_n)E_{n-2} +J_n(\alpha(B_{n-1}+B_n)+\beta )E_{n-1} \nonumber \\
\hphantom{i\frac{{\rm d}}{{\rm d}z}E_n =}{} + \big(\alpha\big(J_n^2+B_n^2 +J_{n+1}^2 \big)+\beta B_n\big)E_n + J_{n+1}(\alpha(B_n+B_{n+1})+\beta)E_{n+1} \nonumber\\
\hphantom{i\frac{{\rm d}}{{\rm d}z}E_n =}{} +\alpha(J_{n+1}J_{n+2})E_{n+2}. \label{nnn}
 \end{gather}
 Here $J_n$ and $B_n$ are the same matrix elements of $J$ as before. The parameters $\alpha$ and $\beta$ determine respectively the strengths of the NNN and NN interactions. The coupling $J_n$ will still be physically realized according to~(\ref{coupling}). In light of~(\ref{nnn}), the NNN couplings will hence be proportional to the distance between the next-to-nearest sites.

 Since $\bar{J}$ has the same eigenbasis as $J$, the condition for PST and FR remain the same except that the spectrum of $\bar{J}$, $\{\alpha\lambda_s^2+\beta\lambda_s\}$ should now be used. The PST condition on the dif\/ference between two successive eigenvalues is therefore
 \begin{gather} \label{2vp}
 (\lambda_{s}-\lambda_{s-1})(\alpha(\lambda_{s}+\lambda_{s-1})+\beta)=\frac{\pi}{Z_{\rm PST}}M_s
 \end{gather}
 and the conditions for FR take the form
 \begin{gather*}
 Z_{\rm FR}(\lambda_{2s}-\lambda_{2s-1})(\alpha(\lambda_{2s}+\lambda_{2s-1})+\beta)=-(-1)^N2\theta+2\pi L_s^{(0)},\\
 Z_{\rm FR}(\lambda_{2s+1}-\lambda_{2s})(\alpha(\lambda_{2s+1}+\lambda_{2s})+\beta)=(-1)^N2\theta+2\pi L_s^{(1)}.
 \end{gather*}
 We shall now spell out what these conditions entail when $J_n$ is given by~(\ref{jkraf}) and $B_n=0$, that is when~$J$ is the Jacobi matrix associated to the Krawtchouk polynomials.
 \subsection{A NNN model based on Krawtchouk couplings}
 \subsubsection{FR}
 The FR conditions require $L_s^{(0)}=c_1s+c_0$ and become, for $s$ even,
 \begin{gather*}
 Z_{\rm FR}(\alpha(4s-(N+1))+\beta) = -(-1)^N2\theta+2\pi(c_1 s +c_0)
 \end{gather*}
 and for $s$ odd with $L_s^{(1)}=c_1's+c_0'$,
 \begin{gather*}
 Z_{\rm FR}(\alpha(4s-(N-1))+\beta) = (-1)^N2\theta+2\pi(c_1' s +c_0').
 \end{gather*}
 These two relations yield the following set of equalities
 \begin{gather*}
 \alpha Z_{\rm FR} = \frac{\pi}{2}c_1, \qquad \alpha Z_{\rm FR} = \frac{\pi}{2}c_1',\\
 Z_{\rm FR}(\beta-\alpha(N+1)) = -(-1)^N2\theta+2\pi c_0,\qquad
 Z_{\rm FR}(\beta-\alpha(N-1)) = (-1)^N2\theta+2\pi c_0',
 \end{gather*}
 which in turn imply $c_1 = c_1'$ and
 \begin{gather*}
 \frac{\beta}{\alpha} = \frac{-(-1)^N2\theta+2\pi c_0}{\pi c_1}+(N+1),\qquad
 \frac{\beta}{\alpha} = \frac{(-1)^N2\theta+2\pi c_0'}{\pi c_1}+(N-1).
 \end{gather*}
 One then arrives at the following formula for the mixing angle $\theta$:
 \begin{gather*}
 \theta = (-1)^N\left(\frac{\pi}{4}c_1+\frac{\pi}{2}(c_0-c_0')\right)
 \end{gather*}
 with the FR distance given by
 \begin{gather*}
 Z_{\rm FR} = \frac{\pi}{2\beta}(c_1 N+2\pi(c_0+c_0')).
 \end{gather*}
 One thus observes that for $N$ odd, balanced FR will happen at distances $\frac{\pi q}{2\beta}$ where~$q$ is odd. For~$N$ even the FR distance will be $\frac{\pi p}{\beta}$ where $p$ is an integer. This is in contradistinction with the situation in the NN Krawtchouk system where FR is not possible.
 \subsubsection{PST}
 Upon substituting $\lambda_s =(s-\frac{N}{2})$ in (\ref{2vp}), with $M_s=2(c_1s+c_0)+1$, one gets
 \begin{gather*}
 \alpha(2s-(N+1))+\beta = \frac{\pi}{Z_{\rm PST}} (2(c_1 s +c_0) +1),
 \end{gather*}
 where $c_1$, $c_0$ are both integers.

 Equating the terms with the same power of $s$, one obtains
 \begin{gather*}
 \alpha Z_{\rm PST} = \pi c_1,\qquad Z_{\rm PST}(\beta-\alpha(N+1)) = \pi(2c_0 +1).
 \end{gather*}
 From these conditions, we see that
 \begin{gather*}
 \frac{\beta}{\alpha} = \frac{2c_0+1}{c_1}+(N+1),
 \end{gather*}
 which indicates that $\frac{\beta}{\alpha}$ must be a rational number. One now f\/inds for $Z_{\rm PST}$:
 \begin{gather*}
 Z_{\rm PST} = \frac{\pi}{\beta} (2c_0+1+c_1(N+1)).
 \end{gather*}
 Since $c_0$ and $c_1$ are integers, only two cases can occur
 \begin{gather*}
1) \quad Z_{\rm PST} = \frac{2\pi j }{\beta},\\
2) \quad Z_{\rm PST} = \frac{\pi }{\beta}(2j+1).
 \end{gather*}
 The f\/irst case only arises when $N$ is even and the second case can materialize for both parities of~$N$. Note that PST will occur at double the FR distance.

 \subsection{Other NNN models}
 The approach to the construction of NNN models using the recurrence coef\/f\/icients of the Krawtchouk polynomials can obviously be extended by considering matrices $J$ associated to other families of orthogonal polynomials. The spectral lattices will then be more involved than the linear one.

 As observed, the analysis of the spectral conditions for FR and PST involves the study of integer-valued polynomials in the integer variable~$s$. For cases beyond the linear one, it is useful to recall that any integer-valued polynomial~$p(s)$ of degree~$N$ in~$s$ can be written in the form
 \begin{gather*}
 p(s) = \sum_{n=0}^{N} c_n \frac{s!}{(n-s)! n!},
 \end{gather*}
 where all the coef\/f\/icients $c_n$ are integers.

 This has proved of help in the examination of other analytic models with NNN couplings. The one based on the dual-Hahn couplings for instance has been conf\/irmed in this way to exhibit FR and PST.

\section{Conclusion}\label{section7}
This paper has of\/fered an overview of many analytic photonic lattices (equivalently of spin chains) with FR and PST and has described their main features. It has shed light on the role played by univariate orthogonal polynomials in the design of these devices. Novel results for an array based on a special case of $q$-Racah polynomials have been obtained, namely that FR can occur when $N$ is odd with the mixing angle depending on the value of $q$.

It is to be expected that the burgeoning theory of multivariate orthogonal polynomials would be instrumental in the construction of higher dimensional simplexes with interesting transport properties. Work in this direction has been initiated \cite{26,27,28}. PST in graphs has also been the object of much attention. The reviews \cite{30, 29} give surveys of this extended body of work. An interesting question would be to study FR in this context. The results in \cite{kay2006perfect} on next-to-nearest neighbour provide a good starting point for numerically determining coupling strengths when long range interactions are taken into account.

Finally, it should prove rewarding to explore evolutions that are more complex than PST and FR. See \cite{31} for steps in this direction.

\subsection*{Acknowledgements}

This survey is based in part on a talk given by one of us (L.V.) at SIDE12. We are grateful to the Guest Editors of the Special Issue in Symmetries and Integrability of Dif\/ference Equations for their invitation to write a topical review. The authors wish to thank Jean-Michel Lemay for his input on the para-Racah model as well as M.~Christandl, V.X.~Genest, H.~Miki, S.~Tsujimoto and A.~Zhedanov for their collaboration on many of the advances presented here. \'E.-O.B.~gratefully acknowledges a scolarship from the Physics Department of Universit\'e de Montr\'eal. The research of L.V.\ is supported in part by a grant from NSERC (Canada). We wish to acknowledge the elaborate and constructive reports of the referees that have helped improve the paper.

\pdfbookmark[1]{References}{ref}
\LastPageEnding

\end{document}